      [1999/12/01 v1.4c Il Nuovo Cimento]
\documentclass{cimento1}

\usepackage{graphicx}  
\title{Standard Model and New Physics: theoretical 
and experimental perspectives\thanks{\small 
Summary talk at `Incontri di Fisica delle Alte Energie',
Bologna, Italy, 26 -- 28 March, 2008.}}

\author{A.~Castro\from{ins:bo}
        \atque
G.~Corcella\from{cf}\from{ins:pi}}
\instlist{\inst{ins:bo} Universit\`a di Bologna and INFN, Sezione di
       Bologna, Italy
\inst{cf} Museo Storico della Fisica e 
Centro Studi e Ricerche E.~Fermi, Roma, Italy
  \inst{ins:pi} Scuola Normale Superiore and INFN, Sezione di Pisa, Italy}
\PACSes{
\PACSit{12.15.-y}{Electroweak interactions}
\PACSit{12.38.-t}{Quantum Chromodynamics}
\PACSit{12.60.-i}{Models beyond the Standard Model}}
\begin{document}

\maketitle

\begin{abstract}
In this summary we present the current status of 
the Standard Model of strong and electroweak interactions 
from the theoretical and experimental point of view. 
Some discussion is also devoted to the 
exploration of possible New Physics signals beyond the Standard Model.
\end{abstract}

\section{Introduction}
The Standard Model is the theory of strong and electroweak interactions: 
they are described in terms 
of quarks and leptons, the basic constituents of matter, and gauge bosons,
which are the carriers of the fundamental forces. 
This model is kept endlessly under inspection by comparing 
measured observables, 
such as couplings, masses, integral and 
differential cross sections or branching ratios, 
with the theory expectations.

The interplay between theory, which is 
continuosly improving its predictions, 
and experiments, whose measurements are 
carried out with increasing accuracy, 
consolidates the Standard Model itself.
At the same time, however, 
the Standard Model presents some drawbacks
which open the road to New Physics; the forthcoming experiments
at the Large Hadron Collider will provide 
us with a unique chance to address these open problems.

We present in the following our perspective, from both 
theoretical and experimental viewpoints, on the current situation of 
the Standard Model and discuss some of its open 
issues, which call for New Physics extensions.

\section{Theoretical and phenomenological issues}

Quantum Chromodynamics (QCD), 
an unbroken renormalizable theory based on the
colour $SU(3)$ gauge symmetry,  
is a main part of the Standard Model of particle physics, 
as it is the theory of strong 
interactions. At present and future colliders, having full control of QCD 
is fundamental, since events mediated by the strong interaction may mimic New Physics signals.

Because of asymptotic freedom, at high 
energies QCD observables can be computed as
fixed-order expansions in the coupling constant $\alpha_S(Q^2)$
(perturbative QCD). Most cross sections 
have been computed at next-to-leading order (NLO), 
whereas a few have been provided with
next-to-next-to-leading order (NNLO) corrections as well.
In fact, as NLO corrections can be as large as a factor of two, 
leading-order (LO)
results can just estimate the order of magnitude of an observable, 
with the NLO giving the first reliable prediction.
NNLO computations 
are nonetheless still necessary to reduce the theoretical uncertainty
and weaken the dependence, {\it e.g.}, on the choice of
parton distributions or fragmentation functions.

Among recent NLO 
calculations, we can mention the corrections to the production of
three vector bosons at the LHC, namely $ZZZ$, $WZZ$, $WWZ$ and $WWW$: 
the results on the transverse momentum $(p_T)$ spectrum show that
the NLO has a relevant impact, indeed up to a factor of two, 
at small $p_T$ \cite{ossola}.

While fixed-order calculations are reliable enough to predict inclusive observables,
such as total cross sections and widths,
differential distributions 
can exhibit contributions where $\alpha_S(Q^2)$ is multiplied
by logarithmic coefficients, which become large for soft or collinear parton
radiation. Resumming such terms is mandatory to describe exclusive quantities.

A remarkable example, 
where the interplay between fixed-order and resummed computation is
crucial, is Higgs-boson production in the $gg\to H$ channel, the dominant 
one at the LHC.
The Higgs rapidity and
transverse momentum spectra have been computed to NNLO accuracy; moreover,
the logarithms $\ln(p_T^2/m_H^2)$, large at small $p_T$, where one is mostly 
sensitive to soft and collinear emissions,
have been resummed in the 
next-to-next-to-leading logarithmic 
(NNLL) approximation and included in the 
Monte Carlo code HqT \cite{bozzi}.

At low energy, however, one is faced with non-perturbative power corrections 
which cannot be calculated in perturbative QCD.
In fact, one may address non-perturbative phenomena, such as hadronisation or
underlying event, following several models and approaches,
which are typically based on fits to experimental data.
As far as hadronisation is concerned, an alternative possibility consists 
in defining an effective strong coupling constant,
free from the Landau pole, 
which at large energy coincides with the standard coupling and at small energy
includes non-perturbative power corrections.
Studies have shown that the effective-coupling 
model, used along with
a resummed calculation in the 
framework of perturbative fragmentation functions,
yields a reasonable description of data on $b$- and
$c$-flavoured hadrons at LEP and SLD, 
without tuning any parameter to such data \cite{ferrera}.

Another approach to non-perturbative QCD parametrises low-energy physics
by means of the
frozen coupling constant, defined as the integral of 
$\alpha_S(p_T^2)$, where $p_T$ is some transverse momentum
characteristic of the process,
from zero to an upper
limit, fitted to experimental data.
In this way, one can, {\it e.g.}, study jet observables at the Tevatron or the LHC,
obtaining results in reasonable agreement with Monte Carlo generators,
such as HERWIG or PYTHIA, 
which implement their own models for hadronisation and
underlying event, 
tuned to LEP or Tevatron data \cite{magnea}.

Along with the intensive applications of QCD to collider phenomenology,
work has been lately undertaken to construct 
a dual model of QCD, in the fashion of the AdS/CFT correspondence.
In the framework of such a model, called holographic QCD, 
one can perform a linear fit of the experimental spectrum
of $\rho$ mesons and give predictions for masses and decay constants
of the $0^{++}$ glueballs \cite{nicotri}.

The electroweak 
sector of the Standard Model is based on a $SU(2)_L\times U(1)_Y$ 
symmetry, spontaneously broken into $U(1)_{\mathrm{em}}$.
Although the electroweak theory has been successfully tested
in several experimental environments, the Higgs boson, a crucial 
ingredient of the model, as it is responsible 
of the mass generation mechanism, has not been discovered yet.
There are, however, 
indications that a mass mechanism in the fashion of the
Higgs one should exist: for instance, the
observed longitudinal polarisations of
$W$ and $Z$ bosons are indeed manifestations of a broken symmetry.
Furthermore, 
the relation between their masses, namely $M_W^2=M_Z^2\cos^2\theta_W$, 
$\theta_W$ being the Weinberg angle, valid up to very small
radiative corrections, confirms the so-called custodial $SU(2)$ symmetry
of the Standard Model and  that 
the Higgs boson has to be a weak-isospin doublet.
 
The LHC has been designed to solve the
Higgs problem, as it will be able to search for a 
Standard Model Higgs boson, with a mass up to $m_H\simeq 1$~TeV/c$^2$. 
On the theory side,  
the lack of the Higgs discovery has opened the road to
other scenarios beyond the Standard Model,
including the possibility that the Higgs may not exist 
(Higgsless models) 
or that it is an approximate Goldstone boson of a broken global
symmetry (composite Higgs models).
Also, conceptual drawbacks, and especially the
well-known hierarchy problem, namely the quadratical divergence 
of the Higgs mass after radiative corrections, requiring 
fine tuning of several orders of magnitude, call for
extensions of the Standard Model. 
Among these, one of the most appealing and widely studied is surely 
supersymmetry, with its minimal formulation, the Minimal
Supersymmetric Standard Model (MSSM).

Calculations and computing codes implementing
supersymmetric processes have been available for a few years.
Work has been lately carried out to provide
such computations with radiative corrections,
which are likely to have a remarkable impact at the
LHC, and must be taken into account in any 
reliable analysis. An example is
given by the inclusion of NLO electroweak corrections to
squark-antisquark pair production at hadron colliders, 
included in the Monte Carlo 
program PROSPINO \cite{mirabella}.

A common feature of several New Physics models is that they
predict the existence of new heavy vector bosons
$Z'$. Indeed, $Z'$ production in
Drell--Yan type processes, followed by leptonic decays,
are among the first New Physics signals possibly visible at the
LHC.
Ref.~\cite{guzzi} discusses $Z'$ production in the so-called 
left-right symmetric free-fermionic model, which enlarges
the gauge structure of the Standard Model by an extra
heterotic-string inspired $U(1)$ (see \cite{guzzi} for more details).
The free-fermionic model is constructed in such a way to
suppress proton decay; moreover, it is
anomaly-free and consistent with the see-saw mechanism for
neutrino masses, with family universality and Yukawa-like couplings.
Within this model, $Z'$-production processes have been 
provided with QCD corrections to 
NNLO accuracy and should be visible at the LHC in the channels
$Z'\to\gamma\gamma$ and $Z'\to ZZ\to \ell_1\bar \ell_1
\ell_2\bar\ell_2$.

Although 
the electroweak precision tests seem to favour a possibly light Higgs,
as will be discussed in detail in the next section,
it is nonetheless still possible to accommodate a heavy Higgs  
beyond the Standard Model, since New Physics effects can  
contribute to electroweak observables
and make a heavy Higgs consistent with present data.
An example is the so-called $\lambda$SUSY model, based on an
extension 
of the MSSM, which predicts the existence of a heavy Higgs $H$, 
with mass, {\it e.g.}, 500-600~GeV/c$^2$, mainly decaying, via $H\to hh$,
into a pair of lighter Higgses $h$, having mass
$m_h\simeq$ 200-300~GeV/c$^2$. 
The properties of the lighter Higgs $h$ 
are supposed to be pretty similar to the Standard Model one.
In fact, 
New Physics contributions, {\it e.g.} to the oblique parameters $S$ and $T$,
make $\lambda$SUSY still compatible with the constraints of
the electroweak precision tests, which
do seem to favour a rather light Higgs, but only within
the Standard Model and without assuming any New Physics effect 
\cite{franceschini}. The $\lambda$SUSY model is expected to be
possibly observable at the LHC with 100~fb$^{-1}$ of integrated
luminosity.

Nevertheless, a possible alternative is to try to describe
the electroweak interactions and solve the hierarchy problem 
in the framework of the so-called Higgsless models.
A process which has been thoroughly investigated is $WW$ scattering:
in the
Standard Model, the exchange of an intermediate $H$ moderates the growth
of the cross section for longitudinal boson scattering, and the requirement
of unitarity sets the limit $m_H<1$~TeV/c$^2$
on the Higgs mass.
In Higgsless models, the starting point is typically
a $SU(2)_L\times SU(2)_R$ chiral symmetry, spontaneously broken to
$SU(2)_{L+R}$:
in this way, one or more new heavy vector bosons replace the
Higgs and delay the unitarity issue to a few TeV/c$^2$.

Ref.~\cite{accomando} discusses the
four-site Higgsless model, which presents four new vector bosons, 
{\it i.e.} two charged $W^\pm_{1,2}$ and two neutral $Z_{1,2}$. 
At the LHC, $W'$ and $Z'$ bosons can be produced
in Drell--Yan like processes, thus
giving rise to
visible resonances.
Thanks to these new gauge bosons, the problem of unitarity violation is 
delayed to energy scales which are much higher than those actually
probed at the LHC. Unlike other Higgsless models, such vector bosons
are not `fermiophobic', as the electroweak precision data do allow 
sizeable couplings with fermions.

Before closing this section, we point out that,
for the sake of experimental analyses, any New Physics model needs 
to be implemented in Monte Carlo generators, 
in order to account for 
multiple initial- and final-state
radiation, hadronisation and underlying event.
In fact, work towards this direction has seen tremendous improvements in the
latest few years. As for $WW$ scattering, it is worthwhile to mention the
PHANTOM code, which implements final states with six fermions,
and includes the option of no intermediate Higgs boson as well.
By using PHANTOM, it will be possible to study $WW$ scattering at the LHC
and, in the phase of 100~fb$^{-1}$ luminosity, discriminate between, {\it e.g.},
a Standard Model scenario with a Higgs of mass $m_H\simeq 200$~GeV/c$^2$ and 
the no-Higgs case \cite{bevilacqua}.

\section{The experimental status}
The Standard Model  continues to be explored in different experiments 
based on accelerators or not. 
The Tevatron, colliding 
beams of protons and antiprotons with the currently highest 
energy of $\sqrt{s}=2$ TeV,  
represents a very favourable environment 
where to perform an extensive study of 
the major ingredients of the Standard Model, 
{\it i.e.} quarks and gauge bosons. 
Figure \ref{tev_proc} shows, as an example, 
the production cross section at the Tevatron for different categories of processes: 
the production of $b$-hadrons,  $W$ and $Z$ bosons, 
events containing the top quark, and finally the possible production of the Higgs boson.  
These and other processes are thouroughly studied by the two 
multi-purpose experiments operating at the Tevatron, namely CDF and D\O. 
\par
Given the cross sections and the current integrated 
luminosity (about 3 fb$^{-1}$), 
the production of events containing $W$ or $Z$ bosons is large, 
of the order of a few hundred thousand events. 
Also, even if the Tevatron cannot be really 
considered a top factory, the production of top quarks is quite abundant, 
with yields, so far,  of about $2\times 10^4$ 
$t\bar t$ pairs and about $6\times 10^3$ single-top events
per year.
On the other hand, 
the production of the Higgs boson is expected to be 
very rare at the Tevatron and only the total integrated luminosity before the final shutdown,
which is foreseen to be about 6-8 fb$^{-1}$, along with 
a full exploration and combination of all possible channels, might 
enable to obtain some results.

\par
The situation taking place at the Tevatron will change drastically  
once the LHC, colliding beams of protons with $\sqrt{s}=14$~TeV,
reaches its full power. 
The production cross sections  at LHC will increase, with respect 
to the Tevatron ones, by  one order of magnitude 
for $W/Z$ production and by two orders for top production. This, 
along with the higher-design 
instantaneous luminosity, 
will make LHC a real top-factory, and a promising environment for Higgs discovery.
\par
Given the different 
yields of the processes described in Figure \ref{tev_proc}, 
it is suitable to divide the experimental studies 
of the major Standard Model processes into three subfields: 
$W/Z$ physics, production of top quarks and searches 
for the Standard Model Higgs boson. 
In addition, we shall briefly
discuss the exploration for New Physics beyond the Standard Model.

\subsection{$W/Z$ physics}
The study of the $W/Z$ boson production, within the Standard Model, 
is crucial because it can provide accurate tests of 
both electroweak and strong interactions; 
QCD corrections to the production cross section are
in fact available in the NNLO approximation.

CDF and D\O~typically measure~\cite{ref:cavaliere} the production 
cross sections for leptonic final states, {\it i.e.}
$\sigma_W\times BR(W\to \ell\nu)$ 
and $\sigma_Z\times BR(Z\to \ell^+\ell^-)$, 
along with their ratio and the $W$-width $\Gamma_W$. 
The size of the samples also
enables the study of differential distributions.
\par
The measurement of the $W$ mass is very important, due 
to its link through radiative electroweak corrections 
to the Higgs boson mass. 
The mass is measured at the Tevatron using the 
large sample of inclusive 
$W$ events, where the $W$ decays into $e\nu$ or $\mu\nu$. 
The $W$ transverse mass, 
$M^T_W=\sqrt{2P_T^\ell P_T^\nu (1-\cos\Delta\Phi^{\ell\nu})}$, 
is reconstructed using the missing transverse 
energy, as a measure of $P_T^\nu$, measuring the azimuthal angle 
$\Delta\Phi^{\ell\nu}$ between the lepton and the 
direction of the missing transverse energy, and is finally fitted 
to the expected distributions for different values of the $W$ mass. 
This measurement is a very delicate 
one, because it relies on a very precise calibration of the lepton momenta, 
and even 
the underlying event and additional $p\bar p$ interactions must
be taken into account. 
With this technique, 
and using only $1/15$ of the luminosity currently available, CDF  
already reaches an accuracy better than what was measured 
by the various experiments at LEP, thus improving the world average.
\par
The study of diboson production $WW$, $WZ$ and 
$ZZ$ is accessible in spite of the 
relatively low production cross sections (respectively about 10,~4 and~1~pb). 
The investigation of these events is quite interesting not only in itself, 
but also because they represent relatively rare processes, 
with topologies similar to what can be expected for Higgs boson production. 
In fact, $WZ$ production 
has been clearly observed and there are first hints of $ZZ$ evidence.
\par
At the LHC, given the high  
energy and luminosity, a large number of $W$ and $Z$ events 
will be collected shortly after the machine turn-on, with a very 
clean leptonic signature. 
Even before 
providing accurate measurements of $\sigma_W$, $\sigma_Z$ and $M_W$, 
and performing other  
tests of the electroweak theory, 
such events will be quite useful for calibration 
and alignment of the detector~\cite{ref:rovelli}. Once the detector 
is well understood, the study at the LHC of diboson production 
will become interesting to search for the Higgs boson and
for deviations from the Standard Model.

\subsection{Top quark physics}
The top quark, discovered thirteen years ago at the Tevatron, 
is the heaviest quark and, for this reason, it 
plays an important role in loop 
corrections to several electroweak observables. 
Indeed, together with the accurate measurement of the $W$ mass, 
the top-mass measurement constrains the mass of the Higgs boson.
\par
The measurement of the top quark mass with the highest precision is one of 
the major goals of CDF and D\O. 
These two experiments already managed to measure the mass 
with a (combined) 1.4  GeV/c$^2$ precision, 
exceeding past expectations~\cite{ref:gresele}. 
The goal now is to reach a precision smaller than 1 GeV/c$^2$, before the 
Tevatron definitive turn-off.
\par
As for the production cross section, 
the uncertainty reached by CDF and D\O~ is of the order of 10\% 
(per experiment) and comparable to the theoretical uncertainties, 
thus enabling a test of NLO/NLL predictions. 
\par
Electroweak production of 
single top has a cross section about 1/3 of $t\bar t$ pair production. 
Therefore, it has been more difficult to find evidence 
for it and to measure the single-top cross section, 
which gives also a direct measurement of the Cabibbo--Kobayashi--Maskawa 
matrix element $V_{tb}$.
\par
In addition to mass and cross sections, other top properties are measured, 
such as its charge, lifetime, 
decay branching ratios and the $W$ helicity in top decays $t\to bW$. 
Also, due to its large mass, top quarks might couple to New Physics at 
high energy scales and act as probes for these processes. 
For this reason,  exotic decays or production mechanisms have been explored, 
but there is so far no evidence of heavy resonances or fourth-generation quarks.
\par
The measurement at the LHC of the top quark properties will reach high 
statistical accuracy, given the large yield expected for top events.  
Precise measurements will require, however, a detailed modelling of 
the detector response, along with a good understanding 
of the sources of systematic uncertainty. 
Top events themselves can be used, 
through {\it in-situ} calibration of the jet energy scale, 
to help reduce such uncertainties~\cite{ref:cobal}. 
\begin{figure}
\begin{minipage}[c]{0.45\linewidth}
\centering
\includegraphics[width=6cm]{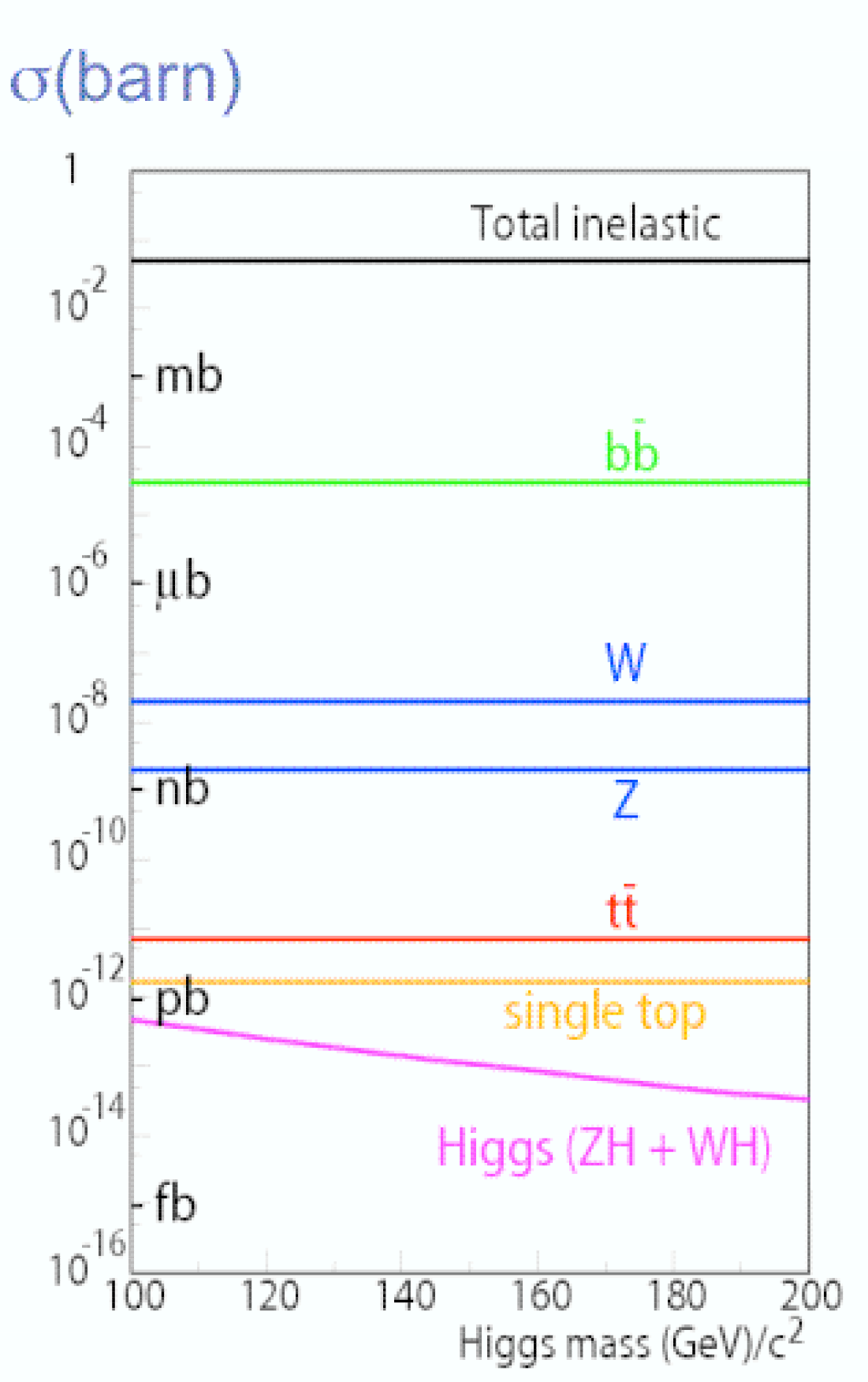}     
\caption{Production cross section at the Tevatron, as a function of the Higgs boson mass, for various processes.}
\label{tev_proc}
 \end{minipage}\hspace{0.5cm}
\begin{minipage}[c]{0.45\linewidth}
\centering
\includegraphics[width=6cm]{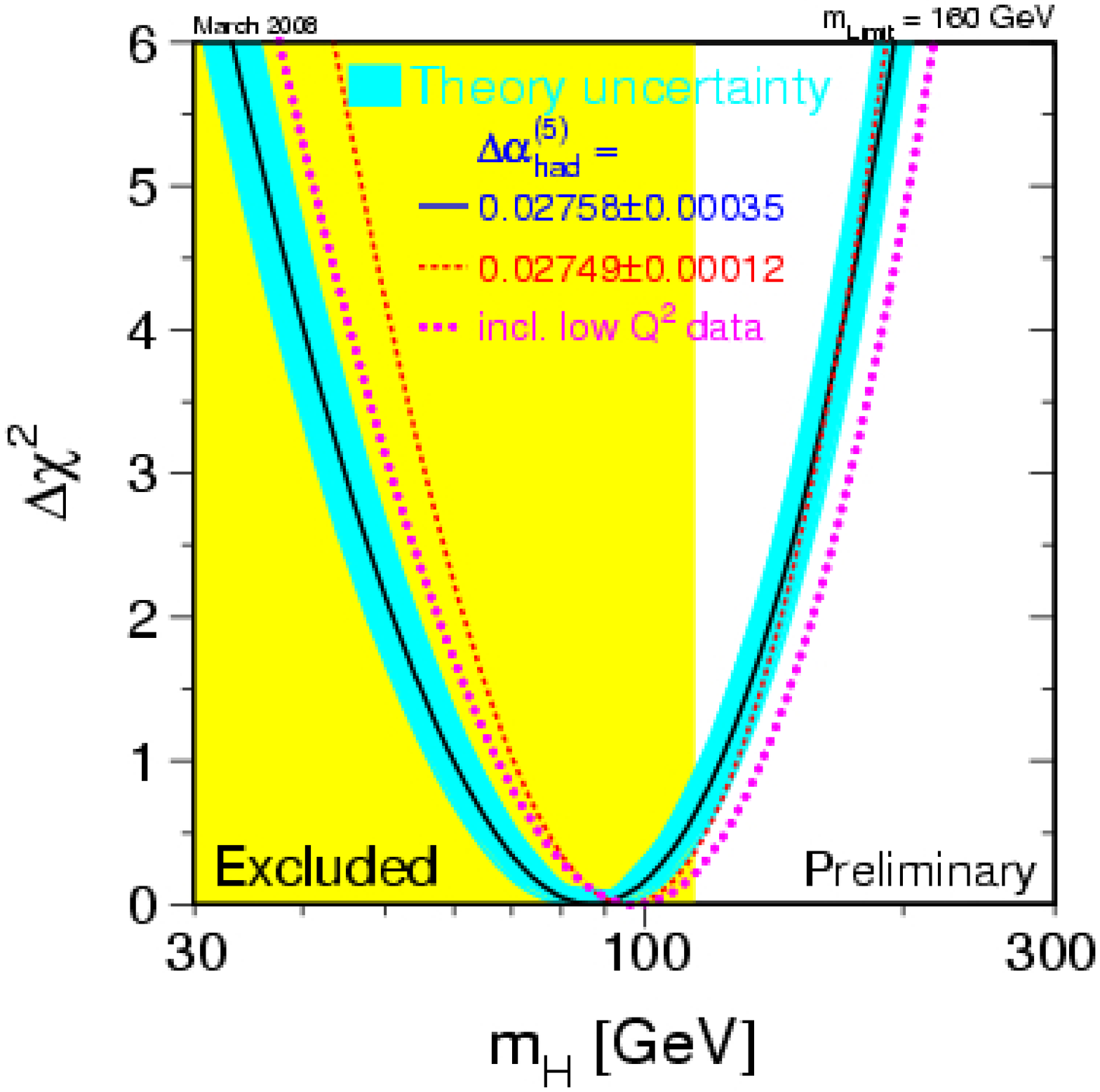}     
\caption{March 2008 $\Delta\chi^2$ curve derived from high-$Q^2$ precision electroweak measurements, performed at LEP and by SLD, CDF, and D\O, as a function of the Higgs boson mass, assuming the Standard Model. The yellow band 
on the left-hand side refers to the LEP2 direct search limit. From ~\cite{ref:LEPEWWG}.}
\label{blue}
\end{minipage}
\end{figure}

\subsection{Standard Model Higgs}
The measurements of $W$ and top masses, together with other 
precision electroweak measurements, 
lead to the limit for the Standard Model 
Higgs-boson mass of $86^{+36}_{-27}$ GeV/c$^2$ (see Figure \ref{blue}), 
with a 95\% C.L. upper limit of 160 GeV/c$^2$. 
If one also includes the results from direct searches at LEP, 
yielding $m_H> 114$~GeV/c$^2$,
the 95\% C.L. limit becomes 190 GeV/c$^2$.
\par\noindent
Such a (relatively low) expected mass sets Higgs-boson production 
possibly within reach of the Tevatron experiments, 
but, at the same time, the smallness 
of the production cross section makes it very hard to search for 
its evidence. 
This challenging task requires the experiments to aim at high sensitivity by 
combining all possible channels, improving efficiencies and reducing systematic uncertainties. 
Specific strategies need to be implemented for a low mass Higgs ($m_H< 135$  GeV/c$^2$), where the associated 
production ($WH$ or $ZH$) has the best chances, and for a (relatively) high mass 
($135\div 200$  GeV/c$^2$), where the channels $H\to WW$ are the most favourable ones.
\par\noindent
The CDF+D\O~ 
sensitivity reaches its optimum around 160 GeV/c$^2$ where they could be able to 
reach some results before the Tevatron turn-off either in terms of exclusion or of evidence~\cite{ref:margaroli}. 
\par
As for CMS and ATLAS, once the detector is well calibrated and its response known, they are expected 
to be sensitive to the
discovery already with $\approx 1$ fb$^{-1}$ of data, 
if the Higgs mass is above 140  
GeV/c$^2$, in the decay channels $H\to ZZ$ and $H\to WW$. For 
lower Higgs masses, more luminosity, 
about 5-10  fb$^{-1}$, will be needed and several
channels will have to be included, in order  
to be sensitive to its
observation~\cite{ref:solfaroli}.

\subsection{New Physics beyond the Standard Model}
As discussed in section 2,
many extensions of the Standard Model are 
introduced in order to address open issues,  
such as the hierarchy problem, or
to provide different mechanisms of symmetry 
breaking or unify the fundamental forces. 
\par
The searches currently performed at the 
Tevatron are either inspired by a specific model 
or instead based on a specific signature.
Searches for supersymmetric particles, 
new gauge bosons, extra dimensions, leptoquarks or Higgs compositeness
belong to the first case.
As for the second case, anomalous production of leptons, 
bosons and photons, as well as missing $E_T$, have been searched.
No signal of supersymmetry, exotic particles 
or anomalous production has yet been found at the 
Tevatron and limits (more or less stringent, 
depending on the cases) have been set~\cite{ref:manca}.
\par
New Physics is searched also at HERA where, 
for instance, limits on contact interactions have been set~\cite{ref:bindi}.
\par
Given the large luminosity that will be integrated, the LHC experiments 
will have the potential for important discoveries, 
either within the SUSY framework~\cite{ref:desanctis} 
or in other scenarios beyond the Standard Model~\cite{ref:bernardini}. 
Statistical uncertainties will not be a problem in most cases, 
but it will be crucial to 
understand the systematics in order to be able to find evidence of new signals.

\section{Conclusions}

In this summary we have briefly reviewed the current 
status of the Standard Model of
strong and electroweak interactions and underlined
some of its drawbacks.
We discussed improvements in the theory 
and their applications to collider phenomenology, as well as the 
most recent measurements 
carried out at present colliders, mostly at the Tevatron accelerator,  
and commented on the features of future measurements which will be
carried out at the LHC.

On the theory side, QCD stands as a robust theory, which
continuously yields predictions which are 
confirmed very well by the experiments.
Nevertheless, more accurate perturbative
calculations and a better understanding and modelling
of its non-perturbative aspects are still of great interests,
in order to fully control the backgrounds for
many New Physics searches.

As for the electroweak interactions, 
the ultimate confirmation of the Standard Model awaits 
the discovery of the Higgs boson: as discussed, the LHC 
should be able to give a final answer to this point. 
Moreover, the absence of any Higgs signal so far, along with 
the open problems of the Standard Model, has pushed the
development of a few new ideas on the 
electroweak symmetry breaking, 
which are theoretically well formulated 
and not in contradiction with the available
electroweak precision tests. 
As only experimental measurements, {\it e.g.}
the $WW$ scattering cross section, can help to
verify or disprove such models, we can
just eagerly await for the next start of the LHC.


\begin{thebibliography}{0}
\bibitem{ossola}\BY{Ossola~G.} these proceedings.
\bibitem{bozzi}\BY{Bozzi~G.} these proceedings.
\bibitem{ferrera}\BY{Ferrera~G.} these proceedings.
\bibitem{magnea}\BY{Magnea~L.} these proceedings.
\bibitem{nicotri}\BY{Nicotri~S.} these proceedings.
\bibitem{mirabella}\BY{Mirabella~E.} these proceedings.
\bibitem{guzzi}\BY{Guzzi~M.} these proceedings.
\bibitem{franceschini}\BY{Franceschini~R.} these proceedings.
\bibitem{accomando}\BY{Accomando~E.} these proceedings.
\bibitem{bevilacqua}\BY{Bevilacqua~G.} these proceedings.
\bibitem{ref:cavaliere}\BY{Cavaliere~V.} these proceedings.
\bibitem{ref:rovelli}\BY{Rovelli~C.} these proceedings.
\bibitem{ref:gresele}\BY{Gresele~A.} these proceedings.
\bibitem{ref:cobal}\BY{Cobal~M.} these proceedings.
\bibitem{ref:LEPEWWG}\BY{The LEP electroweak working group} 
http://lepewwg.web.cern.ch/LEPEWWG/.
\bibitem{ref:margaroli}\BY{Margaroli~F.} these proceedings.
\bibitem{ref:solfaroli}\BY{Solfaroli~E.} these proceedings.
\bibitem{ref:manca}\BY{Manca~G.} these proceedings.
\bibitem{ref:bindi}\BY{Bindi~M.} these proceedings.
\bibitem{ref:desanctis}\BY{De Sanctis~U.} these proceedings.
\bibitem{ref:bernardini}\BY{Bernardini~J.} these proceedings.

\end{thebibliography}
\end{document}